\documentclass[prl,twocolumn,showpacs,superscriptaddress]{revtex4}

\usepackage{amsmath,amssymb}
\usepackage{verbatim}
\usepackage{graphicx}
\usepackage[
      colorlinks=true,
      linkcolor=blue,
      urlcolor=magenta,
      filecolor=green,
      citecolor=red,
      pdfstartview=FitV,
      pdftitle={},
        pdfauthor={Arjun Bagchi, Daniel Grumiller and Wout Merbis},
        pdfsubject={},
        pdfkeywords={},
        pdfpagemode=None,
        bookmarksopen=true
      ]{hyperref}
\usepackage{color}

\DeclareFontFamily{OT1}{rsfs}{}
\DeclareFontShape{OT1}{rsfs}{m}{n}{ <-7> rsfs5 <7-10> rsfs7 <10->rsfs10}{} 
\DeclareMathAlphabet{\mycal}{OT1}{rsfs}{m}{n}

\newcommand{\be}[1]{ \begin{equation}\label{#1} }
\newcommand{\ee}{\end{equation}}
\newcommand{\bea}[1]{\begin{eqnarray}\label{#1} }
\newcommand{\eea}{\end{eqnarray}}

\newcommand{\refb}[1]{(\ref{#1})}

\newcommand{\cL}{{\cal L}}
\newcommand{\cLb}{\bar{\cal L}}

\newcommand{\eq}[2]{\begin{equation} #1 \label{#2} \end{equation}}

\DeclareMathOperator{\extdm}{d}

\newcommand{\extd}{\extdm \!}

\newcommand{\cM}{{\cal M}}
\newcommand{\cN}{{\cal N}}

\newcommand{\beq}{\begin{equation}}
\newcommand{\eeq}{\end{equation}}
\newcommand{\bi}{\begin{itemize}}
\newcommand{\ei}{\end{itemize}}
\newcommand{\bt}{\begin{tabular}}
\newcommand{\et}{\end{tabular}}
\newcommand{\bc}{\begin{center}}
\newcommand{\ec}{\end{center}}

\def\one{{\hbox{ 1\kern-.8mm l}}}
\newcommand{\Dslash}{\not{\hbox{\kern-4pt $D$}}}
\newcommand{\pdslash}{\not{\hbox{\kern-2pt $\partial$}}}

\newcommand{\ti}{N}
\newcommand{\tii}{M}

\newcommand{\ba}{\begin{array}}
\newcommand{\ea}{\end{array}}

\def\bbox{{\,\lower0.9pt\vbox{\hrule \hbox{\vrule height 0.2 cm
\hskip 0.2 cm \vrule height 0.2 cm}\hrule}\,}}
\newcommand{\dsl}{\pa \kern-0.5em /}

\newcommand{\zb}{\bar z}
\newcommand{\partialb}{\bar \partial}

\newcommand{\Ab}{\bar A}

\newcommand{\tr}{\rm Tr}

\begin{document}

%\title{Stress tensor correlators in 3-dimensional gravity and Galilean conformal field theories}
\title{Stress tensor correlators in three-dimensional gravity}

\author{Arjun Bagchi}
\email{abagchi@mit.edu}
\affiliation{Center for Theoretical Physics, Massachusetts Institute of Technology, 77 Massachusetts Avenue, Cambridge, MA 02139-4307, USA.} 
\altaffiliation[On leave of absence from]{ Indian Institute of Science Education and Research, Pune, India}

\author{Daniel Grumiller}
\email{grumil@hep.itp.tuwien.ac.at}
\affiliation{Institute for Theoretical Physics, Vienna University of Technology, Wiedner Hauptstrasse 8--10/136, A-1040 Vienna, Austria}

\author{Wout Merbis}
\email{merbis@hep.itp.tuwien.ac.at}
\affiliation{Institute for Theoretical Physics, Vienna University of Technology, Wiedner Hauptstrasse 8--10/136, A-1040 Vienna, Austria}

\date{\today}

\preprint{TUW--15--13, MIT-CTP--4694}

\begin{abstract} 
We calculate holographically arbitrary $n$-point correlators of the boundary stress tensor in three-dimensional Einstein gravity with negative or vanishing cosmological constant. We provide explicit expressions up to 5-point (connected) correlators and show consistency with the Galilean conformal field theory Ward identities and recursion relations of correlators, which we derive. This provides a novel check of flat space holography in three dimensions.
\end{abstract}


\maketitle

\section{Introduction}

%\paragraph{Introduction.}
Correlation functions are the main observational and theoretical entities in many branches of physics, in particular in any system that effectively is described by some quantum field theory (QFT), with numerous applications in condensed matter physics, cosmology, particle physics and many other research fields. Given their ubi\-quitousness, it is useful to have several tools available to calculate correlators. 

A remarkable tool for strongly coupled QFTs is provided by holography \cite{'tHooft:1993gx,Susskind:1995vu}. In particular, the Anti-de~Sitter/conformal field theory (AdS/CFT) correspondence \cite{Maldacena:1997re,Aharony:1999ti} allows to calculate correlators in a specific strongly coupled CFT by mapping this (typically) hard calculation to a simple one in a specific gravity theory \cite{Gubser:1998bc,Witten:1998qj}. 

As the simplest example, the 1-point function [i.e., the vacuum expectation value (vev)] of the stress tensor $T_{\mu\nu}$ is calculated by taking the first variation of the corresponding gravity action $S_{\textrm{\tiny AdS}}$ with respect to the metric, evaluated on solutions of the equations of motion (EOM):
\eq{
\langle T_{\mu\nu}\rangle_{\textrm{\tiny CFT}} = \frac{\delta S_{\textrm{\tiny AdS}}}{\delta g^{\mu\nu}}\Big|_{\textrm{\tiny EOM}}
}{eq:cor1}
Note that the right hand side is non-zero and finite due to boundary terms \cite{Henningson:1998gx,Balasubramanian:1999re,Emparan:1999pm}. The holographic calculation of arbitrary (connected) $n$-point correlators of the stress tensor is a straightforward generalization of \eqref{eq:cor1} and requires to evaluate the $n^{\textrm{th}}$ variation of the gravity action. The $n$-point correlators of the stress tensor can also be calculated on the CFT side, so ultimately these calculations are a check of the holographic Ward identities. 

For more recent holographic correspondences that go beyond AdS/CFT, such as flat space holography [see \cite{Barnich:2006av,Barnich:2010eb,Bagchi:2010zz, Bagchi:2012cy, Barnich:2012aw,Bagchi:2012yk,Barnich:2012xq,Bagchi:2012xr,Bagchi:2013lma,Afshar:2013vka,Gonzalez:2013oaa,Costa:2013vza,Fareghbal:2013ifa,Bagchi:2013qva, Krishnan:2013wta,Detournay:2014fva,Barnich:2014kra,Barnich:2014cwa,Riegler:2014bia,Fareghbal:2014qga,Bagchi:2014iea,Gary:2014ppa,Oblak:2015sea,Barnich:2015uva,Barnich:2015mui} for some recent advances in three dimensions (3D)], similar checks would contribute to increase the credibility of the purported relations between certain QFTs and their conjectured gravity duals.

The main goal of the present work is to provide the check outlined above for flat space Einstein--Hilbert (EH) gravity in 3D by calculating arbitrary $n$-point correlators of the stress tensor components on the QFT side and, independently, on the gravity side. 
So far only the 0- and 1-point functions were calculated holographically \cite{Bagchi:2013lma,Costa:2013vza,Fareghbal:2013ifa,Detournay:2014fva}. Our results are the first explicit check of the flat space holographic 2- and higher-point correlators and allow to present for the first time explicit results for the 3-, 4- and 5-point correlators of the stress tensor components in two-dimensional (2D) Galilean conformal field theories (GCFTs) as well as their Ward identities.

\section{AdS$\boldsymbol{_3}$/CFT$\boldsymbol{_2}$}

%\paragraph{AdS$_3$/CFT$_2$.}
We start with the AdS$_3$/CFT$_2$ derivation of $n$-point correlators to exhibit the required tools in a familiar context. 
The key result is that all $n$-point correlators of the stress tensor in a CFT$_2$ are determined by the underlying symmetry algebra, which consists of two copies of the Virasoro algebra ($n$, $m$ are integers),
\eq{
[\cL_n,\cL_m] = (n-m)\cL_{n+m} + \frac{c}{12}n(n^2-1)\delta_{n,-m}
}{eq:cor16}
and similarly for the barred generators $\bar\cL_n$, with the central charge $c$ replaced by its barred counterpart $\bar c$.
(The same remark applies to all remaining CFT formulas, so to reduce clutter we display only unbarred quantities.)

In their seminal paper, Belavin, Polyakov and Zamolodchikov proved a recursion relation between $n$- and $(n-1)$-point correlators of the stress tensor in any CFT$_2$ \cite{Belavin:1984vu}. For the connected part of the correlators this recursion relation reads
\eq{
\langle T^1\, T^2 \dots T^n\rangle = \sum_{i=2}^n \bigg(\frac{2}{z_{1i}^2} + \frac{1}{z_{1i}}\,\partial_{z_i}\bigg)\langle T^2 \dots T^n\rangle
}{eq:cor2}
where we introduced the abbreviation $z_{ij}=z_i-z_j$. Here $T^i=T_{zz}(z_i)$ is the holomorphic component of the stress tensor and $z$,~$\bar z$ are coordinates on the plane with metric $g_{zz}=0=g_{\bar z\bar z}$ and $g_{z\bar z}=1$. 
The disconnected contribution to the $n$-point correlators is the sum of all permutations of products of lower-point functions such that all products have $n$ points. Since holographic computations yield connected correlators we focus on them. All correlators in this paper are solely the connected part.

The $n$-point correlators are completely determined by the Ward identities and the 2-point correlators, which in turn are uniquely determined by conformal invariance \eqref{eq:cor16}:
\eq{
\langle T^1\,T^2\rangle = \frac{c}{2z_{12}^4}
}{eq:cor3} 
Using this in the recursion relation \eqref{eq:cor2} gives the higher-point correlators \cite{diFrancesco,Blumenhagen,Osborn:2012vt,Headrick:2015gba}. 
\begin{subequations}
 \label{eq:cor0}
\begin{align}
\langle T^1\,T^2\,T^3\rangle & =  \frac{c}{z_{12}^2z_{23}^2z_{13}^2} \label{eq:cor9} \displaybreak[1] \\
\langle T^1\,T^2\,T^3\,T^4\rangle &=  \frac{2c\,g_4(\gamma)}{z_{14}^2z_{23}^2z_{12}z_{13}z_{24}z_{34}}  \label{eq:cor10} \displaybreak[1] 
\\ \langle T^1\,T^2\,T^3\,T^4\,T^5\rangle &= \frac{4c\, g_5(\gamma,\,\zeta)}{\prod_{1\leq i<j\leq 5} z_{ij}}  \label{eq:cor13}
\end{align}
The functions
\begin{align}
g_4(\gamma) = &\, \frac{\gamma^2-\gamma+1}{\gamma} \label{eq:cor11} 
 \\ g_5 (\gamma,\,\zeta)  = &\, \frac{\gamma+\zeta}{2(\gamma-\zeta)} - \,\frac{(\gamma^2-\gamma\zeta+\zeta^2)}{\gamma(\gamma-1)\zeta(\zeta-1)(\gamma-\zeta)} \times \nonumber \\ 
 & \times \big([\gamma(\gamma-1)+1][\zeta(\zeta-1)+1]-\gamma\zeta\big) \label{eq:cor14}
\end{align}
depend on the cross-ratios
\eq{
\gamma = \frac{z_{12}\,z_{34}}{z_{13}\,z_{24}} \qquad \zeta=\frac{z_{25}\,z_{34}}{z_{35}\,z_{24}}\,.
}{eq:cor12}
\end{subequations}

This procedure can be iterated to arbitrary $n$-point correlators, with increasingly lengthy expressions. 
Thus, if we can holographically establish the recursion relation \eqref{eq:cor2} and calculate the 2-point correlators \eqref{eq:cor3} we have essentially succeeded in holographically computing all the $n$-point correlators of the stress tensor. 
To this end, we consider a deformation of a free CFT action $S_0$ by a source term $\mu$ for the stress tensor,
\eq{
S_\mu = S_0 - \int\extd^2z \,\mu(z,\bar z) T(z)\,.
}{eq:cor6}
We then localize the source at some point $(z_2,\,\bar z_2)$.
\eq{
\mu(z,\bar z) = \epsilon \, \delta^{(2)}(z-z_2,\bar z-\bar z_2)
}{eq:cor7}
Here $\epsilon$ is a small expansion parameter introduced to keep track of the order of $\mu$. The 1-point function with respect to the $\mu$-deformed vacuum yields the 2-point correlator with respect to the free vacuum,
\eq{
\langle T^1\rangle_\mu = \langle T^1\rangle + \epsilon \,\langle T^1\, T^2 \rangle + {\cal O}(\epsilon^2)\,.
}{eq:cor8}

We now compute this expansion in the dual AdS$_3$ gravity theory. According to the AdS/CFT dictionary, the source $\mu$ corresponds to the boundary value of the metric component $g_{\zb\zb}$. To compute the $n$-point function we need the $n^{\textrm{th}}$ variation of the bulk action with respect to the source. Such a computation is lengthy in the metric formulation, but fortunately simplifies significantly in the Chern--Simons (CS) formulation of 3D gravity \cite{Achucarro:1987vz,Witten:1988hc}. In this framework, the EH action is the difference of two CS actions; $I_{\textrm{\tiny EH}}=I_{\textrm{\tiny CS}}[A]-I_{\textrm{\tiny CS}}[\Ab]$ with
\eq{I_{\textrm{\tiny CS}} [A] = \frac{k}{4\pi} \int  \tr \big( A \wedge dA + \tfrac23 A\wedge A \wedge A \big)
}{eq:CSaction}
where $A$ and $\Ab$ are two $sl(2,\mathbb{R})$-valued connections and the CS level $k$ is related to the AdS length $\ell$ and Newton's constant $G_N$ through $k=\frac{\ell}{4G_N}$. We write asymptotically AdS spacetimes in a `highest-weight gauge' \cite{Banados:1998gg}.
\begin{subequations}
 \label{eq:conn}
\begin{align}
A &= b^{-1}(\extd + a)b & b &= e^{\rho L_0} \\
a_z &= L_+ + \frac{\cL}{k} L_- & a_{\zb} &= -\mu L_+ + \ldots  
\end{align}
\end{subequations}
Similar expressions hold for the barred sector. Here $L_{\pm},\,L_0$ are the $sl(2,\mathbb{R})$ generators and $\rho$ is the radial coordinate of AdS. The dots denote terms of lower weight which are completely fixed by the EOM. In this gauge, the `chemical potential' $\mu$ corresponds precisely to the source of the boundary stress tensor and $\cL$ is its vev \cite{Banados:2004nr}, see also \cite{deBoer:2014fra,Li:2015osa}. The EOM are the CFT Ward identities:
\begin{equation}\label{eq:ward}
-\partialb \cL =  \frac{k}{2} \partial^3 \mu + 2 \cL \partial \mu + \mu \partial \cL\,.
\end{equation}
To calculate the 2-point function, we localize the sources as in \eqref{eq:cor7} and expand
$\cL(z)=\cL^{(0)}(z)+\epsilon\cL^{(1)}(z)+{\cal O}(\epsilon^2)$, where
$\cL^{(0)}$ is the background value, which we take to be AdS$_3$ in the Poincar\'e patch $(\cL^{(0)}=\cLb^{(0)}=0)$, while $\cL^{(1)}$ corresponds to the linear term in $\epsilon$ in \eqref{eq:cor8}. This is precisely the 2-point function. From \eqref{eq:ward} we find
\eq{\partialb \cL^{(1)} = - \frac{k}{2}\, \partial^3 \delta^{(2)}(z-z_2,\zb-\zb_2)\,.}{eq:lin}
We solve this differential equation using the Green function of the Laplacean  $\partial\partialb$ in flat space,
$G(z_{12},\,\zb_{12})=\ln\left(z_{12}\zb_{12}\right)$.
Then
\eq{\cL^{(1)} = -\frac{k}{2} \partial_{z_1}^4 G(z_{12}) = \frac{3k}{z_{12}^4}}{eq:hol2pt}
gives exactly the 2-point function \eqref{eq:cor3} with central charge $c=6k=\frac{3\ell}{2G_N}$, compatible with \cite{Brown:1986nw}.

The $n$-point functions can be obtained similarly. We localize the source at $n-1$ points,
$\mu(z,\bar z)=\sum_{i=2}^n\epsilon_i\,\delta^{(2)}(z-z_i,\bar z-\bar z_i):=\sum_{i=2}^n\epsilon_i\delta_i$.
On the CFT side, this gives the expansion
$\langle T^1\rangle_\mu=\langle T^1\rangle+\sum_{i=2}^{n}\epsilon_i\,\langle T^1\,T^i\rangle+\ldots+\big(\prod_{i=2}^n\epsilon_i\big)\langle T^1\,T^2\ldots T^{n}\rangle+{\cal O}(\epsilon^n)$.
The EOM at order $\epsilon^{n-1}$ give
\eq{
-\partialb \cL^{(n-1)} =  \sum_{i=2}^n \epsilon_i \Big( 2 \cL^{(n-2)}  \partial  \delta_i + \delta_i \partial \cL^{(n-2)} \Big)\,. 
}{eq:holnpt1}
Using the Green function $G$ to solve for $\cL^{(n-1)}$ yields
\eq{
\cL^{(n-1)}(z_1) = \sum_{i=2}^n \epsilon_i \left( \frac{2}{z^2_{1i}} + \frac{1}{z_{1i}}\,\partial_{z_i} \right) \cL^{(n-2)}(z_i) \,.
}{eq:recursive}
The sum of terms in this expression proportional to $\Pi_{i=2}^n\epsilon_i$ is the $n$-point function \eqref{eq:cor2}. All other terms are contact terms with two operators localized at the same point. We have now derived holographically the 2-point function and the CFT recursion relation for the correlators \eqref{eq:cor2} in the CS formulation and therefore know that all $n$-point correlators of the stress tensor match between gravity and CFT side.

The results generalize from the plane to the cylinder after taking $z=e^{i\omega}$, with $\omega=it+\varphi$ and $\varphi\sim\varphi+2\pi$, so that $(\omega,\,\bar \omega)\sim(\omega+2\pi,\,\bar\omega+2\pi)$. In all correlators effectively the only change is that one needs to replace the quantities $z_{ij}\to 2\,\sin{[(\omega_i-\omega_j)/2]}=:s_{ij}$ and \eqref{eq:cor2} becomes ($c_{ij}:=\textrm{cot}[(\omega_i-\omega_j)/2]$)
\eq{
 \langle T^1\, T^2 \dots T^n\rangle_{\rm C} = \sum_{i=2}^n \Big(\frac{2}{s_{1i}^2}+\frac{c_{1i}}{2}\,\partial_{\omega_i}\Big)\langle T^2 \dots T^n\rangle_{\rm C}
}{eq:whatever}
On the gravity side the relevant vacuum state is now global AdS$_3$ with background values ${\cal L}^{(0)}=\bar{\cal L}^{(0)}=\frac{k}{4}$.

\section{GCFT stress tensor correlators}

%\paragraph{GCFT stress tensor correlators.}
The asymptotic symmetries of 3D flat space at null infinity are governed by the BMS$_3$ algebra \cite{Ashtekar:1996cd,Barnich:2006av} whose non-vanishing commutators are ($n,\,m$ are integers)
\begin{subequations}
 \label{eq:angelinajolie}
\begin{align}
 [L_n,L_m] &= (n-m)L_{n+m} + \frac{c_L}{12}n(n^2-1)\delta_{n,-m} \\
 [L_n,M_m] &= (n-m)M_{n+m} + \frac{c_M}{12}n(n^2-1)\delta_{n,-m} \end{align}
\end{subequations}
This algebra also arises out of non- or ultra-relativistic contractions of two copies of the Virasoro algebra \eqref{eq:cor16} and is known as 2D Galilean Conformal Algebra (GCA). QFTs with GCA symmetries are called GCFTs. If there exists a dual formulation of 3D flat space gravity in terms of a 2D QFT it has to be a GCFT$_2$ \cite{Barnich:2006av,Bagchi:2010zz, Bagchi:2012cy}. 

The stress tensor components $\cM_{\text{P}},\,\cN_{\text{P}}$ of a
GCFT$_2$ on the plane can be mode-expanded in terms of $L_n,\,M_n$
\cite{Bagchi:2010vw, Bagchi:2013bga},
$\cM_{\text{P}}=\sum_nM_n\xi^{-n-2},\,\cN_{\text{P}}=\sum_n \, \big(L_n-(n-2)\frac{\eta}{\xi} M_n \big)\xi^{-n-2}$.
The mapping from the plane to the cylinder is given by
$\xi=e^{i\varphi},\,\eta=iue^{i\varphi}$, with $\varphi\sim\varphi+2\pi$.
On the cylinder one has the following mode expansions \cite{Bagchi:2013bga}:
\begin{subequations}
\label{emc}
\begin{align}
-M + \frac{c_M}{24} &= \xi^2 \cM_{\text{P}}  = \sum_n M_n e^{-in \varphi}\\
-N + \frac{c_L}{24} &=  \xi^2 \cN_{\text{P}} - 2 \xi \eta \cM_{\text{P}}  = \sum_n (L_n - i n u M_n) e^{-in \varphi} 
\end{align}
\end{subequations}
For our explorations of flat holography we work on the cylinder and consider highest-weight representations of the GCFT$_2$ \cite{Bagchi:2009ca, Bagchi:2009pe}. We define the vacuum of the theory $|0\rangle$ as 
$L_n|0\rangle=0=M_n|0\rangle,\,\forall n\geq-1$.
Using the above relation and its conjugate ($\langle 0| L_n  = 0 = \langle 0| M_n, \ n\leq 1$), together with \refb{emc} and \refb{eq:angelinajolie}, we can calculate any correlation function of the GCFT$_2$ stress tensor components. In particular, the non-vanishing 2-point functions in the cylinder representation are given by
\eq{
\langle \tii^1\, \ti^2 \rangle = \frac{c_M}{2 s_{12}^4}  \qquad 
\langle \ti^1\, \ti^2 \rangle = \frac{c_L - 2c_M\tau_{12}}{2 s_{12}^4}
}{eq:fc0} 
with the definitions, $M^i=M(\varphi_i),\,N^i=N(u_i,\,\varphi_i)$, $s_{ij}:=2\sin{[(\varphi_i-\varphi_j)/2]}$, $c_{ij}:=\cot{[(\varphi_i-\varphi_j)/2]}$ and $\tau_{ij}:=(u_i-u_j)\,c_{ij}$.

Using the mode expansion and the algebra, the higher-point correlators can be derived similarly. 
Note that (the connected part of) any correlator with two or more $M$-insertions vanishes.
The first few non-vanishing higher-point functions read
\begin{subequations}
 \label{eq:cor95}
\begin{align} 
 &\langle \tii^1\, \ti^2\, \ti^3 \rangle = \frac{c_M}{s^2_{12}s^2_{13}s^2_{23}} \label{eq:cor100}  \displaybreak[1] \\ 
 &\langle \ti^1\, \ti^2\, \ti^3 \rangle = \frac{c_L-c_M\,\tau_{123}}{s^2_{12}s^2_{13}s^2_{23}}  \displaybreak[1] \\
 &\langle \tii^1\, \ti^2\, \ti^3\,  \ti^4\rangle = \frac{2c_M\, g_4(\gamma)}{s^2_{14} s^2_{23} s_{12} s_{13}s_{24} s_{34}}  \displaybreak[1] \\
 &\langle \ti^1\, \ti^2\, \ti^3\, \ti^4\rangle = \frac{2c_L\,g_4(\gamma)+c_M \Delta_4}{s^2_{14} s^2_{23} s_{12} s_{13}s_{24} s_{34}}  \displaybreak[1] \\
 & \langle \tii^1 \,\ti^2 \,\ti^3\, \ti^4\, \ti^5\rangle = \frac{4c_M\, g_5(\gamma,\,\zeta)}{\prod_{1\leq i < j \leq 5}s_{ij}}  \displaybreak[1] \\
 & \langle \ti^1 \,\ti^2 \,\ti^3 \,\ti^4\, \ti^5\rangle = \frac{4c_L\, g_5(\gamma,\,\zeta)+c_M \Delta_5}{\prod_{1\leq i < j \leq 5}s_{ij}}  \label{eq:cor101}
\end{align}
\end{subequations}
where $\Delta_4 = 4 g^\prime_4(\gamma)\eta_{1234} - (\tau_{1234}+\tau_{14}+\tau_{23})g_4(\gamma)$ and $\Delta_5 = 4 \partial_\gamma g_5(\gamma,\zeta)\eta_{1234} + 4 \partial_\zeta g_5(\gamma,\zeta)\eta_{2345} - 2g_5(\gamma,\zeta) \tau_{12345}$.
In the above, we have defined $\tau_{1\dots n}=\sum_{1\leq i<j\leq n}\tau_{ij}$, e.g.~$\tau_{123}=\tau_{12}+\tau_{13}+\tau_{23}$.
The function $g_4(\gamma)$ [$g_5(\gamma,\zeta)$] is again given by \eqref{eq:cor11} [\eqref{eq:cor14}], where $z_{ij}$ has to be replaced by $s_{ij}$ in the cross-ratios \eqref{eq:cor12}. The quantity $\eta_{pqrs}$ is defined as
$\eta_{pqrs}=\sum^\prime(-1)^{1+i-j}(u_i-u_j)\sin(\varphi_k-\varphi_l)/(s^2_{pr}s^2_{qs})$, and the sum $\sum^\prime$ goes over all six permutations of $\{i,j,k,l\}=\pi(p,q,r,s)$ where $i<j$ and $k<l$.

\section{Recursion relations}

%\paragraph{Recursion relations.} 
We derive now GCFT recursion relations involving the stress tensor components that encode the corresponding Ward identities \cite{Bagchi:2009pe}. The GCFT stress tensor components on the cylinder, \eqref{emc}, obey the conservation equations 
\eq{
\partial_u\langle \tii(\varphi) {\cal O} \rangle = 0 \qquad \partial_u\langle \ti(u,\,\varphi) {\cal O}  \rangle = \partial_\varphi\langle \tii(\varphi) {\cal O}  \rangle 
}{eq:cor99}
for any operator (or product of operators) ${\cal O}$. The left equation \eqref{eq:cor99} looks very similar to the CFT conservation equation $\partial_{\bar z}\langle T(z)\, {\cal O}\rangle=0$ \cite{Belavin:1984vu}, which together with the similarity of the GCA \eqref{eq:angelinajolie} to the Virasoro algebra \eqref{eq:cor16} eventually leads to a recursion relation similar to \eqref{eq:whatever}.
\eq{
\langle \tii^1\, \ti^2 \dots \ti^n\rangle = \sum_{i=2}^n \Big(\frac{2}{s_{1i}^2} + \frac{c_{1i}}{2}  \partial_{\varphi_i} \Big) \,\langle\tii^2\,\ti^3\dots\ti^n\rangle
}{eq:re1}
The right equation \eqref{eq:cor99} allows to determine the correlation functions with only $\ti$-insertions in terms of correlations functions with one $\tii$-insertion \eqref{eq:re1}.
\begin{multline}
\langle \ti^1 \ti^2 \dots \ti^n\rangle = \frac{c_L}{c_M} \langle \tii^1 \ti^2 \ldots \ti^n \rangle 
\\
 + \sum_{i=1}^n u_i \partial_{\varphi_i} \langle \tii^1 \ti^2 \ldots \ti^n \rangle
\label{eq:re2}
\end{multline}
The second recursion relation \eqref{eq:re2} can be derived as follows.
Integrating the right equation \eqref{eq:cor99} and using cyclic symmetry of the $\langle\tii\,\ti\dots\ti\rangle$ correlators we obtain $\langle\ti^1\dots\ti^n\rangle=A+\sum_{i=1}^nu_i\partial_{\varphi_i}\langle\tii^1\,\ti^2\dots\ti^n\rangle$. The quantity $A$ is a $u$-independent integration constant that must be identical to the result in a chiral CFT with central charge $c_L$, since the $u$-independent part of $\ti$ contains only the Virasoro generators $L_n$. As the $\langle\tii\,\ti\dots\ti\rangle$ correlators also give chiral CFT results, with central charge $c_M$, we can write $A=c_L/c_M\langle\tii^1\,\ti^2\dots\ti^n\rangle$. 

We have checked that the recursion relations \eqref{eq:re1}, \eqref{eq:re2} also follow from a suitable contraction of CFT results \eqref{eq:whatever} and that they are compatible with all our explicit results for correlators, \eqref{eq:fc0}-\eqref{eq:cor101}. Thus we have succeeded in deriving the GCFT$_2$ analog of the CFT recursion relation on the cylinder \eqref{eq:whatever}. In the remainder of the paper we derive the same results holographically.

\section{Flat space correlators}

%\paragraph{Flat space correlators.}
The 3D flat space EH action can be rewritten as the Chern-Simons action \eqref{eq:CSaction} for the $iso(2,1)$ connection
$A=A_M^nM_n+A_L^nL_n$, 
where $L_n,\,M_n$ ($n=0,\pm1$) generate the global part of the GCA \eqref{eq:angelinajolie} and the trace is defined as ${\tr}(L_mM_n)=-2\gamma_{mn}$, where $\gamma=\textrm{antidiag}(1,-\frac12,1)$, see \cite{Gary:2014ppa} for our conventions, details and references. Just like in the AdS case, we parametrize the radial dependence of $A$ by going to a gauge where \cite{Barnich:2014cwa,Gary:2014ppa}
$A=b^{-1}(\extd+a)b$
with the group element $b=\exp(\tfrac{r}{2}M_-)$ and $a=a_u\extd u+a_{\varphi}\extd\varphi$ with $u,r,\varphi$ the (outgoing) Eddington--Finkelstein coordinates. In the absence of chemical potentials the connection components take the form
 $a_u=M_+ + \frac{\cal M}{2k}\,M_-$,
 $a_\varphi=L_+ + \frac{\cal M}{2k}\,L_- + \frac{\cal N}{2k}\,M_-$.
Here the CS-level $k$ is related to Newton's constant as $k=1/(4G_N)$. The EOM restrict the state-dependent functions ${\cal M},\,{\cal N}$ as follows: $\partial_u{\cal M}=0,\,\partial_u{\cal N}=\partial_\varphi{\cal M}$. These equations are equivalent to the GCFT conservation equations \eqref{eq:cor99}.

We proceed as in the AdS case and introduce chemical potentials as sources. The component $a_\varphi$ remains unaffected, while $a_u$ changes to $a_u+\Delta a_u$ with
$\Delta a_u =-\mu_MM_+-\mu_LL_++a_0M_0+a_1L_0+a_2M_-+a_3L_-$.
The EOM determine completely the functions $a_i$ in terms of the chemical potentials $\mu_M$ and $\mu_L$ and restrict the state-dependent functions as follows \cite{Gary:2014ppa}.
\begin{subequations}
 \label{eq:EOM}
\begin{align}
-\partial_u \cM & =  k \partial_{\varphi}^3 \mu_L + \mu_L \partial_\varphi \cM + 2 \cM \partial_\varphi \mu_L \displaybreak[1]
\\
-\partial_u \cN & =  k \partial_{\varphi}^3 \mu_M - (1-\mu_M) \partial_{\varphi}  \cM + 2\cM \partial_\varphi \mu_M \nonumber \\
&\quad + \mu_L \partial_\varphi \cN + 2 \cN \partial_{\varphi}\mu_L
\end{align}
\end{subequations}
Analogous to the AdS case, these equations can be interpreted as GCFT Ward identities.

For our background we choose global Minkowski space, ${\cal M}^{(0)}= k/2$, ${\cal N}^{(0)}=0$. To calculate 2-point correlators in this Minkowski background we switch on chemical potentials, localized at $(u_2,\,\varphi_2)$, like in the AdS case \eqref{eq:cor7},
$\mu_{M/L}=\epsilon_{M/L}\,\delta^{(2)}(u-u_2,\,\varphi-\varphi_2):=\epsilon_{M/L}\,\delta$.  
Here $\epsilon_{M/L}$ are constants. Plugging these localized chemical potentials 
into the EOM \eqref{eq:EOM} with ${\cal M}=k/2+{\cal M}^{(1)}$, ${\cal N}={\cal N}^{(1)}$ yields the linearized Ward identities.
\begin{subequations}
 \label{eq:cor96}
\begin{align}
\partial_u \cM^{(1)}  = &- k\, \epsilon_L \big(\partial^3_\varphi \delta + \partial_\varphi \delta \big) 
\label{eq:lalapetz0} \\
\partial_u \cN^{(1)}  = & - k\, \epsilon_M \big(\partial^3_\varphi \delta  + \partial_\varphi \delta \big) +  \partial_{\varphi}  \cM^{(1)} 
\label{eq:lalapetz}
\end{align}
\end{subequations}
To solve them we use the Green function defined as 
$\partial_u \partial_\varphi G(u-u_2,\,\varphi-\varphi_2) = \delta$.
Using the method of images yields
$G(u-u_2,\,\varphi-\varphi_2)=\ln\big((u-u_2)\sin[(\varphi-\varphi_2)/2]\big)$
and thus we can solve the linearized Ward identities \eqref{eq:lalapetz0},\,\eqref{eq:lalapetz},
\eq{
\cM^{(1)}  =  \frac{6 k \epsilon_L}{s_{12}^4} \qquad
\cN^{(1)}  =  \frac{6k(\epsilon_M-2\epsilon_L\,\tau_{12})}{ s_{12}^4}
}{eq:cor97}
where $s_{ij}$ and $\tau_{ij}$ are defined below \eqref{eq:fc0}. The $\epsilon_L$ component of $\cM^{(1)}$ (and the $\epsilon_M$ component of $\cN^{(1)}$) is the 2-point function $\langle \tii^1\, \ti^2 \rangle$ and the $\epsilon_L$ part of $\cN^{(1)}$ is the 2-point function $\langle \ti^1\, \ti^2 \rangle$. They agree precisely with the GCFT 2-point functions \eqref{eq:fc0} for $c_L=0$ and $c_M= 12k =3/G_N$, compatible with \cite{Barnich:2006av,Bagchi:2012yk}.

Like in the AdS case we compute the higher-point functions by localizing the sources at multiple points $u_i,\varphi_i$ and expanding the vev's of the deformed theory to order $\epsilon^{n-1}$, which eventually establishes recursive formulas.
\begin{subequations}
 \label{eq:hore}
 \begin{align}
  \cM^{(n-1)}  &=  \sum_{i=2}^{n} \epsilon_{L}^i\Big(\frac{2}{s_{1i}^2} + \frac{c_{1i}}{2}\partial_{\varphi_i} \Big) \cM^{(n-2)} \label{eq:hore1} \\
  \cN^{(n-1)} &=  \sum_{i=2}^{n} \!\bigg\{\!\Big(\frac{2}{s_{1i}^2} + \frac{c_{1i}}{2}\partial_{\varphi_i} \Big) \big(\epsilon_{M}^i\cM^{(n-2)} + \epsilon_L^i\cN^{(n-2)}\big) \nonumber \\
+ u_i \epsilon_L^i \Big(&\frac{2c_{1i}}{s^2_{ij}} + \frac{\partial_{\varphi_i}}{s_{1i}^2} \Big) \cM^{(n-2)}  \bigg\} + u_1 \partial_{\varphi_1}\cM^{(n-1)}  \label{eq:hore2}
\end{align}
\end{subequations}
The $\epsilon_L$ component of $\cN^{(n-1)}$ then corresponds to the $n$-point function $\langle \ti^1 \,\ti^2 \ldots \ti^n \rangle$. The $\epsilon_M$ component of $\cN^{(n-1)}$ corresponds to the $n$-point correlator with one $\tii^i$ insertion (just as the $\epsilon_{L}$ component of $\cM^{(n-1)}$).
The recursion relation \eqref{eq:hore1} then simply gives \eqref{eq:re1} with $c_M=12k$, while the $\epsilon_L$ part of \eqref{eq:hore2} can be simplified to give
\eq{ \langle \ti^1 \,\ti^2 \ldots \ti^n \rangle = \sum_{i=1}^{n} u_i \partial_{\varphi_i} \langle \tii^1 \,\ti^2 \ldots \ti^n \rangle \,. }{eq:nptFS}
These results show precise agreement with the GCFT recursion relations \eqref{eq:re1}, \eqref{eq:re2}, provided that $c_L = 0$ and $c_M = 12k = 3/G_N$, again compatible with \cite{Barnich:2006av,Bagchi:2012yk}.

\section{Conclusions}

%\paragraph{Conclusions.}
Since the holographic recursion relations \eqref{eq:hore},~\eqref{eq:nptFS} coincide with the ones that we found in a GCFT \eqref{eq:re1},~\eqref{eq:re2} with the EH values of the central charges, $c_L=0,\,c_M=3/G_N$, and the 2-point correlators coincide as well, we conclude that flat space holography is perfectly consistent with the GCFT Ward identities. In other words, all holographic $n$-point correlators of the stress tensor components match precisely with the corresponding GCFT (connected) correlators. 

It would be interesting to generalize our analysis to other 3D gravity (or gravity-like) theories, see e.g.~Refs.~\cite{Deser:1982vy,Deser:1982sw,Achucarro:1987vz,Bergshoeff:2009hq,Andringa:2009yc,Bergshoeff:2013xma,Henneaux:2010xg,Campoleoni:2010zq,Afshar:2013vka,Gonzalez:2013oaa,Gary:2012ms,Henneaux:2012ny} for various intriguing models.

\section{Acknowledgments}

\acknowledgments

We are grateful to Stefan Stricker for discussions.

AB was supported by the Fulbright Foundation. DG was supported by the FWF projects Y~435-N16, I~952-N16 and I~1030-N27.
WM was supported by the FWF project P 27182-N27.

%\bibliographystyle{apsrev4-1}
%\bibliography{review}

\begin{thebibliography}{60}%
\makeatletter
\providecommand \@ifxundefined [1]{%
 \@ifx{#1\undefined}
}%
\providecommand \@ifnum [1]{%
 \ifnum #1\expandafter \@firstoftwo
 \else \expandafter \@secondoftwo
 \fi
}%
\providecommand \@ifx [1]{%
 \ifx #1\expandafter \@firstoftwo
 \else \expandafter \@secondoftwo
 \fi
}%
\providecommand \natexlab [1]{#1}%
\providecommand \enquote  [1]{``#1''}%
\providecommand \bibnamefont  [1]{#1}%
\providecommand \bibfnamefont [1]{#1}%
\providecommand \citenamefont [1]{#1}%
\providecommand \href@noop [0]{\@secondoftwo}%
\providecommand \href [0]{\begingroup \@sanitize@url \@href}%
\providecommand \@href[1]{\@@startlink{#1}\@@href}%
\providecommand \@@href[1]{\endgroup#1\@@endlink}%
\providecommand \@sanitize@url [0]{\catcode `\\12\catcode `\$12\catcode
  `\&12\catcode `\#12\catcode `\^12\catcode `\_12\catcode `\%12\relax}%
\providecommand \@@startlink[1]{}%
\providecommand \@@endlink[0]{}%
\providecommand \url  [0]{\begingroup\@sanitize@url \@url }%
\providecommand \@url [1]{\endgroup\@href {#1}{\urlprefix }}%
\providecommand \urlprefix  [0]{URL }%
\providecommand \Eprint [0]{\href }%
\providecommand \doibase [0]{http://dx.doi.org/}%
\providecommand \selectlanguage [0]{\@gobble}%
\providecommand \bibinfo  [0]{\@secondoftwo}%
\providecommand \bibfield  [0]{\@secondoftwo}%
\providecommand \translation [1]{[#1]}%
\providecommand \BibitemOpen [0]{}%
\providecommand \bibitemStop [0]{}%
\providecommand \bibitemNoStop [0]{.\EOS\space}%
\providecommand \EOS [0]{\spacefactor3000\relax}%
\providecommand \BibitemShut  [1]{\csname bibitem#1\endcsname}%
\let\auto@bib@innerbib\@empty
%</preamble>
\bibitem [{\citenamefont {'t~Hooft}(1993)}]{'tHooft:1993gx}%
  \BibitemOpen
  \bibfield  {author} {\bibinfo {author} {\bibfnamefont {G.}~\bibnamefont
  {'t~Hooft}},\ }in\ \href@noop {} {\emph {\bibinfo {booktitle}
  {Salamfestschrift}}}\ (\bibinfo  {publisher} {World Scientific},\ \bibinfo
  {year} {1993})\ \Eprint {http://arxiv.org/abs/gr-qc/9310026} {gr-qc/9310026}
  \BibitemShut {NoStop}%
%%CITATION = GR-QC 9310026;%%
\bibitem [{\citenamefont {Susskind}(1995)}]{Susskind:1995vu}%
  \BibitemOpen
  \bibfield  {author} {\bibinfo {author} {\bibfnamefont {L.}~\bibnamefont
  {Susskind}},\ }\href@noop {} {\bibfield  {journal} {\bibinfo  {journal} {J.
  Math. Phys.}\ }\textbf {\bibinfo {volume} {36}},\ \bibinfo {pages} {6377}
  (\bibinfo {year} {1995})},\ \Eprint {http://arxiv.org/abs/hep-th/9409089}
  {hep-th/9409089} \BibitemShut {NoStop}%
%%CITATION = HEP-TH 9409089;%%
\bibitem [{\citenamefont {Maldacena}(1998)}]{Maldacena:1997re}%
  \BibitemOpen
  \bibfield  {author} {\bibinfo {author} {\bibfnamefont {J.~M.}\ \bibnamefont
  {Maldacena}},\ }\href@noop {} {\bibfield  {journal} {\bibinfo  {journal}
  {Adv. Theor. Math. Phys.}\ }\textbf {\bibinfo {volume} {2}},\ \bibinfo
  {pages} {231} (\bibinfo {year} {1998})},\ \Eprint
  {http://arxiv.org/abs/hep-th/9711200} {hep-th/9711200} \BibitemShut {NoStop}%
%%CITATION = HEP-TH 9711200;%%
\bibitem [{\citenamefont {Aharony}\ \emph {et~al.}(2000)\citenamefont
  {Aharony}, \citenamefont {Gubser}, \citenamefont {Maldacena}, \citenamefont
  {Ooguri},\ and\ \citenamefont {Oz}}]{Aharony:1999ti}%
  \BibitemOpen
  \bibfield  {author} {\bibinfo {author} {\bibfnamefont {O.}~\bibnamefont
  {Aharony}}, \bibinfo {author} {\bibfnamefont {S.~S.}\ \bibnamefont {Gubser}},
  \bibinfo {author} {\bibfnamefont {J.~M.}\ \bibnamefont {Maldacena}}, \bibinfo
  {author} {\bibfnamefont {H.}~\bibnamefont {Ooguri}}, \ and\ \bibinfo {author}
  {\bibfnamefont {Y.}~\bibnamefont {Oz}},\ }\href@noop {} {\bibfield  {journal}
  {\bibinfo  {journal} {Phys. Rept.}\ }\textbf {\bibinfo {volume} {323}},\
  \bibinfo {pages} {183} (\bibinfo {year} {2000})},\ \Eprint
  {http://arxiv.org/abs/hep-th/9905111} {hep-th/9905111} \BibitemShut {NoStop}%
%%CITATION = HEP-TH 9905111;%%
\bibitem [{\citenamefont {Gubser}\ \emph {et~al.}(1998)\citenamefont {Gubser},
  \citenamefont {Klebanov},\ and\ \citenamefont {Polyakov}}]{Gubser:1998bc}%
  \BibitemOpen
  \bibfield  {author} {\bibinfo {author} {\bibfnamefont {S.~S.}\ \bibnamefont
  {Gubser}}, \bibinfo {author} {\bibfnamefont {I.~R.}\ \bibnamefont
  {Klebanov}}, \ and\ \bibinfo {author} {\bibfnamefont {A.~M.}\ \bibnamefont
  {Polyakov}},\ }\href@noop {} {\bibfield  {journal} {\bibinfo  {journal}
  {Phys. Lett.}\ }\textbf {\bibinfo {volume} {B428}},\ \bibinfo {pages} {105}
  (\bibinfo {year} {1998})},\ \Eprint {http://arxiv.org/abs/hep-th/9802109}
  {hep-th/9802109} \BibitemShut {NoStop}%
%%CITATION = HEP-TH 9802109;%%
\bibitem [{\citenamefont {Witten}(1998)}]{Witten:1998qj}%
  \BibitemOpen
  \bibfield  {author} {\bibinfo {author} {\bibfnamefont {E.}~\bibnamefont
  {Witten}},\ }\href@noop {} {\bibfield  {journal} {\bibinfo  {journal} {Adv.
  Theor. Math. Phys.}\ }\textbf {\bibinfo {volume} {2}},\ \bibinfo {pages}
  {253} (\bibinfo {year} {1998})},\ \Eprint
  {http://arxiv.org/abs/hep-th/9802150} {hep-th/9802150} \BibitemShut {NoStop}%
%%CITATION = HEP-TH 9802150;%%
\bibitem [{\citenamefont {Henningson}\ and\ \citenamefont
  {Skenderis}(1998)}]{Henningson:1998gx}%
  \BibitemOpen
  \bibfield  {author} {\bibinfo {author} {\bibfnamefont {M.}~\bibnamefont
  {Henningson}}\ and\ \bibinfo {author} {\bibfnamefont {K.}~\bibnamefont
  {Skenderis}},\ }\href@noop {} {\bibfield  {journal} {\bibinfo  {journal}
  {JHEP}\ }\textbf {\bibinfo {volume} {07}},\ \bibinfo {pages} {023} (\bibinfo
  {year} {1998})},\ \Eprint {http://arxiv.org/abs/hep-th/9806087}
  {hep-th/9806087} \BibitemShut {NoStop}%
%%CITATION = HEP-TH 9806087;%%
\bibitem [{\citenamefont {Balasubramanian}\ and\ \citenamefont
  {Kraus}(1999)}]{Balasubramanian:1999re}%
  \BibitemOpen
  \bibfield  {author} {\bibinfo {author} {\bibfnamefont {V.}~\bibnamefont
  {Balasubramanian}}\ and\ \bibinfo {author} {\bibfnamefont {P.}~\bibnamefont
  {Kraus}},\ }\href@noop {} {\bibfield  {journal} {\bibinfo  {journal} {Commun.
  Math. Phys.}\ }\textbf {\bibinfo {volume} {208}},\ \bibinfo {pages} {413}
  (\bibinfo {year} {1999})},\ \Eprint {http://arxiv.org/abs/hep-th/9902121}
  {hep-th/9902121} \BibitemShut {NoStop}%
%%CITATION = HEP-TH 9902121;%%
\bibitem [{\citenamefont {Emparan}\ \emph {et~al.}(1999)\citenamefont
  {Emparan}, \citenamefont {Johnson},\ and\ \citenamefont
  {Myers}}]{Emparan:1999pm}%
  \BibitemOpen
  \bibfield  {author} {\bibinfo {author} {\bibfnamefont {R.}~\bibnamefont
  {Emparan}}, \bibinfo {author} {\bibfnamefont {C.~V.}\ \bibnamefont
  {Johnson}}, \ and\ \bibinfo {author} {\bibfnamefont {R.~C.}\ \bibnamefont
  {Myers}},\ }\href@noop {} {\bibfield  {journal} {\bibinfo  {journal} {Phys.
  Rev.}\ }\textbf {\bibinfo {volume} {D60}},\ \bibinfo {pages} {104001}
  (\bibinfo {year} {1999})},\ \Eprint {http://arxiv.org/abs/hep-th/9903238}
  {hep-th/9903238} \BibitemShut {NoStop}%
%%CITATION = HEP-TH 9903238;%%
\bibitem [{\citenamefont {Barnich}\ and\ \citenamefont
  {Comp{\`e}re}(2007)}]{Barnich:2006av}%
  \BibitemOpen
  \bibfield  {author} {\bibinfo {author} {\bibfnamefont {G.}~\bibnamefont
  {Barnich}}\ and\ \bibinfo {author} {\bibfnamefont {G.}~\bibnamefont
  {Comp{\`e}re}},\ }\href {\doibase 10.1088/0264-9381/24/5/F01,
  10.1088/0264-9381/24/11/C01} {\bibfield  {journal} {\bibinfo  {journal}
  {Class.Quant.Grav.}\ }\textbf {\bibinfo {volume} {24}},\ \bibinfo {pages}
  {F15} (\bibinfo {year} {2007})},\ \Eprint
  {http://arxiv.org/abs/gr-qc/0610130} {arXiv:gr-qc/0610130 [gr-qc]}
  \BibitemShut {NoStop}%
%%CITATION = GR-QC/0610130;%%
\bibitem [{\citenamefont {Barnich}\ and\ \citenamefont
  {Troessaert}(2010)}]{Barnich:2010eb}%
  \BibitemOpen
  \bibfield  {author} {\bibinfo {author} {\bibfnamefont {G.}~\bibnamefont
  {Barnich}}\ and\ \bibinfo {author} {\bibfnamefont {C.}~\bibnamefont
  {Troessaert}},\ }\href {\doibase 10.1007/JHEP05(2010)062} {\bibfield
  {journal} {\bibinfo  {journal} {JHEP}\ }\textbf {\bibinfo {volume} {1005}},\
  \bibinfo {pages} {062} (\bibinfo {year} {2010})},\ \Eprint
  {http://arxiv.org/abs/1001.1541} {arXiv:1001.1541 [hep-th]} \BibitemShut
  {NoStop}%
%%CITATION = ARXIV:1001.1541;%%
\bibitem [{\citenamefont {Bagchi}(2010)}]{Bagchi:2010zz}%
  \BibitemOpen
  \bibfield  {author} {\bibinfo {author} {\bibfnamefont {A.}~\bibnamefont
  {Bagchi}},\ }\href {\doibase 10.1103/PhysRevLett.105.171601} {\bibfield
  {journal} {\bibinfo  {journal} {Phys.Rev.Lett.}\ }\textbf {\bibinfo {volume}
  {105}},\ \bibinfo {pages} {171601} (\bibinfo {year} {2010})}\BibitemShut
  {NoStop}%
%%CITATION = PRLTA,105,171601;%%
\bibitem [{\citenamefont {Bagchi}\ and\ \citenamefont
  {Fareghbal}(2012)}]{Bagchi:2012cy}%
  \BibitemOpen
  \bibfield  {author} {\bibinfo {author} {\bibfnamefont {A.}~\bibnamefont
  {Bagchi}}\ and\ \bibinfo {author} {\bibfnamefont {R.}~\bibnamefont
  {Fareghbal}},\ }\href {\doibase 10.1007/JHEP10(2012)092} {\bibfield
  {journal} {\bibinfo  {journal} {JHEP}\ }\textbf {\bibinfo {volume} {1210}},\
  \bibinfo {pages} {092} (\bibinfo {year} {2012})},\ \Eprint
  {http://arxiv.org/abs/1203.5795} {arXiv:1203.5795 [hep-th]} \BibitemShut
  {NoStop}%
%%CITATION = ARXIV:1203.5795;%%
\bibitem [{\citenamefont {Barnich}\ \emph {et~al.}(2012)\citenamefont
  {Barnich}, \citenamefont {Gomberoff},\ and\ \citenamefont
  {Gonzalez}}]{Barnich:2012aw}%
  \BibitemOpen
  \bibfield  {author} {\bibinfo {author} {\bibfnamefont {G.}~\bibnamefont
  {Barnich}}, \bibinfo {author} {\bibfnamefont {A.}~\bibnamefont {Gomberoff}},
  \ and\ \bibinfo {author} {\bibfnamefont {H.~A.}\ \bibnamefont {Gonzalez}},\
  }\href {\doibase 10.1103/PhysRevD.86.024020} {\bibfield  {journal} {\bibinfo
  {journal} {Phys.Rev.}\ }\textbf {\bibinfo {volume} {D86}},\ \bibinfo {pages}
  {024020} (\bibinfo {year} {2012})},\ \Eprint {http://arxiv.org/abs/1204.3288}
  {arXiv:1204.3288 [gr-qc]} \BibitemShut {NoStop}%
%%CITATION = ARXIV:1204.3288;%%
\bibitem [{\citenamefont {Bagchi}\ \emph {et~al.}(2012)\citenamefont {Bagchi},
  \citenamefont {Detournay},\ and\ \citenamefont {Grumiller}}]{Bagchi:2012yk}%
  \BibitemOpen
  \bibfield  {author} {\bibinfo {author} {\bibfnamefont {A.}~\bibnamefont
  {Bagchi}}, \bibinfo {author} {\bibfnamefont {S.}~\bibnamefont {Detournay}}, \
  and\ \bibinfo {author} {\bibfnamefont {D.}~\bibnamefont {Grumiller}},\ }\href
  {\doibase 10.1103/PhysRevLett.109.151301} {\bibfield  {journal} {\bibinfo
  {journal} {Phys.Rev.Lett.}\ }\textbf {\bibinfo {volume} {109}},\ \bibinfo
  {pages} {151301} (\bibinfo {year} {2012})},\ \Eprint
  {http://arxiv.org/abs/1208.1658} {arXiv:1208.1658 [hep-th]} \BibitemShut
  {NoStop}%
%%CITATION = ARXIV:1208.1658;%%
\bibitem [{\citenamefont {Barnich}(2012)}]{Barnich:2012xq}%
  \BibitemOpen
  \bibfield  {author} {\bibinfo {author} {\bibfnamefont {G.}~\bibnamefont
  {Barnich}},\ }\href {\doibase 10.1007/JHEP10(2012)095} {\bibfield  {journal}
  {\bibinfo  {journal} {JHEP}\ }\textbf {\bibinfo {volume} {1210}},\ \bibinfo
  {pages} {095} (\bibinfo {year} {2012})},\ \Eprint
  {http://arxiv.org/abs/1208.4371} {arXiv:1208.4371 [hep-th]} \BibitemShut
  {NoStop}%
%%CITATION = ARXIV:1208.4371;%%
\bibitem [{\citenamefont {Bagchi}\ \emph
  {et~al.}(2013{\natexlab{a}})\citenamefont {Bagchi}, \citenamefont
  {Detournay}, \citenamefont {Fareghbal},\ and\ \citenamefont
  {Simon}}]{Bagchi:2012xr}%
  \BibitemOpen
  \bibfield  {author} {\bibinfo {author} {\bibfnamefont {A.}~\bibnamefont
  {Bagchi}}, \bibinfo {author} {\bibfnamefont {S.}~\bibnamefont {Detournay}},
  \bibinfo {author} {\bibfnamefont {R.}~\bibnamefont {Fareghbal}}, \ and\
  \bibinfo {author} {\bibfnamefont {J.}~\bibnamefont {Simon}},\ }\href
  {\doibase 10.1103/PhysRevLett.110.141302} {\bibfield  {journal} {\bibinfo
  {journal} {Phys. Rev. Lett.}\ }\textbf {\bibinfo {volume} {110}},\ \bibinfo
  {pages} {141302} (\bibinfo {year} {2013}{\natexlab{a}})},\ \Eprint
  {http://arxiv.org/abs/1208.4372} {arXiv:1208.4372 [hep-th]} \BibitemShut
  {NoStop}%
%%CITATION = ARXIV:1208.4372;%%
\bibitem [{\citenamefont {Bagchi}\ \emph
  {et~al.}(2013{\natexlab{b}})\citenamefont {Bagchi}, \citenamefont
  {Detournay}, \citenamefont {Grumiller},\ and\ \citenamefont
  {Simon}}]{Bagchi:2013lma}%
  \BibitemOpen
  \bibfield  {author} {\bibinfo {author} {\bibfnamefont {A.}~\bibnamefont
  {Bagchi}}, \bibinfo {author} {\bibfnamefont {S.}~\bibnamefont {Detournay}},
  \bibinfo {author} {\bibfnamefont {D.}~\bibnamefont {Grumiller}}, \ and\
  \bibinfo {author} {\bibfnamefont {J.}~\bibnamefont {Simon}},\ }\href
  {\doibase 10.1103/PhysRevLett.111.181301} {\bibfield  {journal} {\bibinfo
  {journal} {Phys.Rev.Lett.}\ }\textbf {\bibinfo {volume} {111}},\ \bibinfo
  {pages} {181301} (\bibinfo {year} {2013}{\natexlab{b}})},\ \Eprint
  {http://arxiv.org/abs/1305.2919} {arXiv:1305.2919 [hep-th]} \BibitemShut
  {NoStop}%
%%CITATION = ARXIV:1305.2919;%%
\bibitem [{\citenamefont {Afshar}\ \emph {et~al.}(2013)\citenamefont {Afshar},
  \citenamefont {Bagchi}, \citenamefont {Fareghbal}, \citenamefont
  {Grumiller},\ and\ \citenamefont {Rosseel}}]{Afshar:2013vka}%
  \BibitemOpen
  \bibfield  {author} {\bibinfo {author} {\bibfnamefont {H.}~\bibnamefont
  {Afshar}}, \bibinfo {author} {\bibfnamefont {A.}~\bibnamefont {Bagchi}},
  \bibinfo {author} {\bibfnamefont {R.}~\bibnamefont {Fareghbal}}, \bibinfo
  {author} {\bibfnamefont {D.}~\bibnamefont {Grumiller}}, \ and\ \bibinfo
  {author} {\bibfnamefont {J.}~\bibnamefont {Rosseel}},\ }\href {\doibase
  10.1103/PhysRevLett.111.121603} {\bibfield  {journal} {\bibinfo  {journal}
  {Phys.Rev.Lett.}\ }\textbf {\bibinfo {volume} {111}},\ \bibinfo {pages}
  {121603} (\bibinfo {year} {2013})},\ \Eprint {http://arxiv.org/abs/1307.4768}
  {arXiv:1307.4768 [hep-th]} \BibitemShut {NoStop}%
%%CITATION = ARXIV:1307.4768;%%
\bibitem [{\citenamefont {Gonzalez}\ \emph {et~al.}(2013)\citenamefont
  {Gonzalez}, \citenamefont {Matulich}, \citenamefont {Pino},\ and\
  \citenamefont {Troncoso}}]{Gonzalez:2013oaa}%
  \BibitemOpen
  \bibfield  {author} {\bibinfo {author} {\bibfnamefont {H.~A.}\ \bibnamefont
  {Gonzalez}}, \bibinfo {author} {\bibfnamefont {J.}~\bibnamefont {Matulich}},
  \bibinfo {author} {\bibfnamefont {M.}~\bibnamefont {Pino}}, \ and\ \bibinfo
  {author} {\bibfnamefont {R.}~\bibnamefont {Troncoso}},\ }\href {\doibase
  10.1007/JHEP09(2013)016} {\bibfield  {journal} {\bibinfo  {journal} {JHEP}\
  }\textbf {\bibinfo {volume} {1309}},\ \bibinfo {pages} {016} (\bibinfo {year}
  {2013})},\ \Eprint {http://arxiv.org/abs/1307.5651} {arXiv:1307.5651
  [hep-th]} \BibitemShut {NoStop}%
%%CITATION = ARXIV:1307.5651;%%
\bibitem [{\citenamefont {Caldeira~Costa}(2014)}]{Costa:2013vza}%
  \BibitemOpen
  \bibfield  {author} {\bibinfo {author} {\bibfnamefont {R.}~\bibnamefont
  {Caldeira~Costa}},\ }\href {\doibase 10.1103/PhysRevD.90.104018} {\bibfield
  {journal} {\bibinfo  {journal} {Phys.Rev.}\ }\textbf {\bibinfo {volume}
  {D90}},\ \bibinfo {pages} {104018} (\bibinfo {year} {2014})},\ \Eprint
  {http://arxiv.org/abs/1311.7339} {arXiv:1311.7339 [hep-th]} \BibitemShut
  {NoStop}%
%%CITATION = ARXIV:1311.7339;%%
\bibitem [{\citenamefont {Fareghbal}\ and\ \citenamefont
  {Naseh}(2014)}]{Fareghbal:2013ifa}%
  \BibitemOpen
  \bibfield  {author} {\bibinfo {author} {\bibfnamefont {R.}~\bibnamefont
  {Fareghbal}}\ and\ \bibinfo {author} {\bibfnamefont {A.}~\bibnamefont
  {Naseh}},\ }\href {\doibase 10.1007/JHEP03(2014)005} {\bibfield  {journal}
  {\bibinfo  {journal} {JHEP}\ }\textbf {\bibinfo {volume} {1403}},\ \bibinfo
  {pages} {005} (\bibinfo {year} {2014})},\ \Eprint
  {http://arxiv.org/abs/1312.2109} {arXiv:1312.2109 [hep-th]} \BibitemShut
  {NoStop}%
%%CITATION = ARXIV:1312.2109;%%
\bibitem [{\citenamefont {Bagchi}\ and\ \citenamefont
  {Basu}(2014)}]{Bagchi:2013qva}%
  \BibitemOpen
  \bibfield  {author} {\bibinfo {author} {\bibfnamefont {A.}~\bibnamefont
  {Bagchi}}\ and\ \bibinfo {author} {\bibfnamefont {R.}~\bibnamefont {Basu}},\
  }\href {\doibase 10.1007/JHEP03(2014)020} {\bibfield  {journal} {\bibinfo
  {journal} {JHEP}\ }\textbf {\bibinfo {volume} {1403}},\ \bibinfo {pages}
  {020} (\bibinfo {year} {2014})},\ \Eprint {http://arxiv.org/abs/1312.5748}
  {arXiv:1312.5748 [hep-th]} \BibitemShut {NoStop}%
%%CITATION = ARXIV:1312.5748;%%
\bibitem [{\citenamefont {Krishnan}\ \emph {et~al.}(2014)\citenamefont
  {Krishnan}, \citenamefont {Raju},\ and\ \citenamefont
  {Roy}}]{Krishnan:2013wta}%
  \BibitemOpen
  \bibfield  {author} {\bibinfo {author} {\bibfnamefont {C.}~\bibnamefont
  {Krishnan}}, \bibinfo {author} {\bibfnamefont {A.}~\bibnamefont {Raju}}, \
  and\ \bibinfo {author} {\bibfnamefont {S.}~\bibnamefont {Roy}},\ }\href
  {\doibase 10.1007/JHEP03(2014)036} {\bibfield  {journal} {\bibinfo  {journal}
  {JHEP}\ }\textbf {\bibinfo {volume} {1403}},\ \bibinfo {pages} {036}
  (\bibinfo {year} {2014})},\ \Eprint {http://arxiv.org/abs/1312.2941}
  {arXiv:1312.2941 [hep-th]} \BibitemShut {NoStop}%
%%CITATION = ARXIV:1312.2941;%%
\bibitem [{\citenamefont {Detournay}\ \emph {et~al.}(2014)\citenamefont
  {Detournay}, \citenamefont {Grumiller}, \citenamefont {Sch{\"o}ller},\ and\
  \citenamefont {Simon}}]{Detournay:2014fva}%
  \BibitemOpen
  \bibfield  {author} {\bibinfo {author} {\bibfnamefont {S.}~\bibnamefont
  {Detournay}}, \bibinfo {author} {\bibfnamefont {D.}~\bibnamefont
  {Grumiller}}, \bibinfo {author} {\bibfnamefont {F.}~\bibnamefont
  {Sch{\"o}ller}}, \ and\ \bibinfo {author} {\bibfnamefont {J.}~\bibnamefont
  {Simon}},\ }\href {\doibase 10.1103/PhysRevD.89.084061} {\bibfield  {journal}
  {\bibinfo  {journal} {Phys.Rev.}\ }\textbf {\bibinfo {volume} {D89}},\
  \bibinfo {pages} {084061} (\bibinfo {year} {2014})},\ \Eprint
  {http://arxiv.org/abs/1402.3687} {arXiv:1402.3687 [hep-th]} \BibitemShut
  {NoStop}%
%%CITATION = ARXIV:1402.3687;%%
\bibitem [{\citenamefont {Barnich}\ and\ \citenamefont
  {Oblak}(2014)}]{Barnich:2014kra}%
  \BibitemOpen
  \bibfield  {author} {\bibinfo {author} {\bibfnamefont {G.}~\bibnamefont
  {Barnich}}\ and\ \bibinfo {author} {\bibfnamefont {B.}~\bibnamefont
  {Oblak}},\ }\href {\doibase 10.1007/JHEP06(2014)129} {\bibfield  {journal}
  {\bibinfo  {journal} {JHEP}\ }\textbf {\bibinfo {volume} {1406}},\ \bibinfo
  {pages} {129} (\bibinfo {year} {2014})},\ \Eprint
  {http://arxiv.org/abs/1403.5803} {arXiv:1403.5803 [hep-th]} \BibitemShut
  {NoStop}%
%%CITATION = ARXIV:1403.5803;%%
\bibitem [{\citenamefont {Barnich}\ \emph {et~al.}(2014)\citenamefont
  {Barnich}, \citenamefont {Donnay}, \citenamefont {Matulich},\ and\
  \citenamefont {Troncoso}}]{Barnich:2014cwa}%
  \BibitemOpen
  \bibfield  {author} {\bibinfo {author} {\bibfnamefont {G.}~\bibnamefont
  {Barnich}}, \bibinfo {author} {\bibfnamefont {L.}~\bibnamefont {Donnay}},
  \bibinfo {author} {\bibfnamefont {J.}~\bibnamefont {Matulich}}, \ and\
  \bibinfo {author} {\bibfnamefont {R.}~\bibnamefont {Troncoso}},\ }\href
  {\doibase 10.1007/JHEP08(2014)071} {\bibfield  {journal} {\bibinfo  {journal}
  {JHEP}\ }\textbf {\bibinfo {volume} {1408}},\ \bibinfo {pages} {071}
  (\bibinfo {year} {2014})},\ \Eprint {http://arxiv.org/abs/1407.4275}
  {arXiv:1407.4275 [hep-th]} \BibitemShut {NoStop}%
%%CITATION = ARXIV:1407.4275;%%
\bibitem [{\citenamefont {Riegler}(2015)}]{Riegler:2014bia}%
  \BibitemOpen
  \bibfield  {author} {\bibinfo {author} {\bibfnamefont {M.}~\bibnamefont
  {Riegler}},\ }\href {\doibase 10.1103/PhysRevD.91.024044} {\bibfield
  {journal} {\bibinfo  {journal} {Phys.Rev.}\ }\textbf {\bibinfo {volume}
  {D91}},\ \bibinfo {pages} {024044} (\bibinfo {year} {2015})},\ \Eprint
  {http://arxiv.org/abs/1408.6931} {arXiv:1408.6931 [hep-th]} \BibitemShut
  {NoStop}%
%%CITATION = ARXIV:1408.6931;%%
\bibitem [{\citenamefont {Fareghbal}\ and\ \citenamefont
  {Naseh}(2015)}]{Fareghbal:2014qga}%
  \BibitemOpen
  \bibfield  {author} {\bibinfo {author} {\bibfnamefont {R.}~\bibnamefont
  {Fareghbal}}\ and\ \bibinfo {author} {\bibfnamefont {A.}~\bibnamefont
  {Naseh}},\ }\href {\doibase 10.1088/0264-9381/32/13/135013} {\bibfield
  {journal} {\bibinfo  {journal} {Class.Quant.Grav.}\ }\textbf {\bibinfo
  {volume} {32}},\ \bibinfo {pages} {135013} (\bibinfo {year} {2015})},\
  \Eprint {http://arxiv.org/abs/1408.6932} {arXiv:1408.6932 [hep-th]}
  \BibitemShut {NoStop}%
%%CITATION = ARXIV:1408.6932;%%
\bibitem [{\citenamefont {Bagchi}\ \emph {et~al.}(2015)\citenamefont {Bagchi},
  \citenamefont {Basu}, \citenamefont {Grumiller},\ and\ \citenamefont
  {Riegler}}]{Bagchi:2014iea}%
  \BibitemOpen
  \bibfield  {author} {\bibinfo {author} {\bibfnamefont {A.}~\bibnamefont
  {Bagchi}}, \bibinfo {author} {\bibfnamefont {R.}~\bibnamefont {Basu}},
  \bibinfo {author} {\bibfnamefont {D.}~\bibnamefont {Grumiller}}, \ and\
  \bibinfo {author} {\bibfnamefont {M.}~\bibnamefont {Riegler}},\ }\href
  {\doibase 10.1103/PhysRevLett.114.111602} {\bibfield  {journal} {\bibinfo
  {journal} {Phys.Rev.Lett.}\ }\textbf {\bibinfo {volume} {114}},\ \bibinfo
  {pages} {111602} (\bibinfo {year} {2015})},\ \Eprint
  {http://arxiv.org/abs/1410.4089} {arXiv:1410.4089 [hep-th]} \BibitemShut
  {NoStop}%
%%CITATION = ARXIV:1410.4089;%%
\bibitem [{\citenamefont {Gary}\ \emph {et~al.}(2015)\citenamefont {Gary},
  \citenamefont {Grumiller}, \citenamefont {Riegler},\ and\ \citenamefont
  {Rosseel}}]{Gary:2014ppa}%
  \BibitemOpen
  \bibfield  {author} {\bibinfo {author} {\bibfnamefont {M.}~\bibnamefont
  {Gary}}, \bibinfo {author} {\bibfnamefont {D.}~\bibnamefont {Grumiller}},
  \bibinfo {author} {\bibfnamefont {M.}~\bibnamefont {Riegler}}, \ and\
  \bibinfo {author} {\bibfnamefont {J.}~\bibnamefont {Rosseel}},\ }\href
  {\doibase 10.1007/JHEP01(2015)152} {\bibfield  {journal} {\bibinfo  {journal}
  {JHEP}\ }\textbf {\bibinfo {volume} {1501}},\ \bibinfo {pages} {152}
  (\bibinfo {year} {2015})},\ \Eprint {http://arxiv.org/abs/1411.3728}
  {arXiv:1411.3728 [hep-th]} \BibitemShut {NoStop}%
%%CITATION = ARXIV:1411.3728;%%
\bibitem [{\citenamefont {Oblak}(2015)}]{Oblak:2015sea}%
  \BibitemOpen
  \bibfield  {author} {\bibinfo {author} {\bibfnamefont {B.}~\bibnamefont
  {Oblak}},\ }\href@noop {} {\  (\bibinfo {year} {2015})},\ \Eprint
  {http://arxiv.org/abs/1502.03108} {arXiv:1502.03108 [hep-th]} \BibitemShut
  {NoStop}%
%%CITATION = ARXIV:1502.03108;%%
\bibitem [{\citenamefont {Barnich}\ and\ \citenamefont
  {Oblak}(2015)}]{Barnich:2015uva}%
  \BibitemOpen
  \bibfield  {author} {\bibinfo {author} {\bibfnamefont {G.}~\bibnamefont
  {Barnich}}\ and\ \bibinfo {author} {\bibfnamefont {B.}~\bibnamefont
  {Oblak}},\ }\href {\doibase 10.1007/JHEP03(2015)033} {\bibfield  {journal}
  {\bibinfo  {journal} {JHEP}\ }\textbf {\bibinfo {volume} {1503}},\ \bibinfo
  {pages} {033} (\bibinfo {year} {2015})},\ \Eprint
  {http://arxiv.org/abs/1502.00010} {arXiv:1502.00010 [hep-th]} \BibitemShut
  {NoStop}%
%%CITATION = ARXIV:1502.00010;%%
\bibitem [{\citenamefont {Barnich}\ \emph {et~al.}(2015)\citenamefont
  {Barnich}, \citenamefont {Gonzalez}, \citenamefont {Maloney},\ and\
  \citenamefont {Oblak}}]{Barnich:2015mui}%
  \BibitemOpen
  \bibfield  {author} {\bibinfo {author} {\bibfnamefont {G.}~\bibnamefont
  {Barnich}}, \bibinfo {author} {\bibfnamefont {H.~A.}\ \bibnamefont
  {Gonzalez}}, \bibinfo {author} {\bibfnamefont {A.}~\bibnamefont {Maloney}}, \
  and\ \bibinfo {author} {\bibfnamefont {B.}~\bibnamefont {Oblak}},\ }\href
  {\doibase 10.1007/JHEP04(2015)178} {\bibfield  {journal} {\bibinfo  {journal}
  {JHEP}\ }\textbf {\bibinfo {volume} {1504}},\ \bibinfo {pages} {178}
  (\bibinfo {year} {2015})},\ \Eprint {http://arxiv.org/abs/1502.06185}
  {arXiv:1502.06185 [hep-th]} \BibitemShut {NoStop}%
%%CITATION = ARXIV:1502.06185;%%
\bibitem [{\citenamefont {Belavin}\ \emph {et~al.}(1984)\citenamefont
  {Belavin}, \citenamefont {Polyakov},\ and\ \citenamefont
  {Zamolodchikov}}]{Belavin:1984vu}%
  \BibitemOpen
  \bibfield  {author} {\bibinfo {author} {\bibfnamefont {A.}~\bibnamefont
  {Belavin}}, \bibinfo {author} {\bibfnamefont {A.~M.}\ \bibnamefont
  {Polyakov}}, \ and\ \bibinfo {author} {\bibfnamefont {A.}~\bibnamefont
  {Zamolodchikov}},\ }\href {\doibase 10.1016/0550-3213(84)90052-X} {\bibfield
  {journal} {\bibinfo  {journal} {Nucl.Phys.}\ }\textbf {\bibinfo {volume}
  {B241}},\ \bibinfo {pages} {333} (\bibinfo {year} {1984})}\BibitemShut
  {NoStop}%
%%CITATION = NUPHA,B241,333;%%
\bibitem [{\citenamefont {Di~Francesco}\ \emph {et~al.}(1997)\citenamefont
  {Di~Francesco}, \citenamefont {Mathieu},\ and\ \citenamefont
  {Senechal}}]{diFrancesco}%
  \BibitemOpen
  \bibfield  {author} {\bibinfo {author} {\bibfnamefont {P.}~\bibnamefont
  {Di~Francesco}}, \bibinfo {author} {\bibfnamefont {P.}~\bibnamefont
  {Mathieu}}, \ and\ \bibinfo {author} {\bibfnamefont {D.}~\bibnamefont
  {Senechal}},\ }\href@noop {} {\emph {\bibinfo {title} {Conformal Field
  Theory}}}\ (\bibinfo  {publisher} {Springer},\ \bibinfo {year}
  {1997})\BibitemShut {NoStop}%
\bibitem [{\citenamefont {Blumenhagen}\ and\ \citenamefont
  {Plauschinn}(2009)}]{Blumenhagen}%
  \BibitemOpen
  \bibfield  {author} {\bibinfo {author} {\bibfnamefont {R.}~\bibnamefont
  {Blumenhagen}}\ and\ \bibinfo {author} {\bibfnamefont {E.}~\bibnamefont
  {Plauschinn}},\ }\href@noop {} {\emph {\bibinfo {title} {Introduction to
  conformal field theory with applications to string theory}}},\ Lecture Notes
  in Physics 779\ (\bibinfo  {publisher} {Springer},\ \bibinfo {year}
  {2009})\BibitemShut {NoStop}%
\bibitem [{\citenamefont {Osborn}(2012)}]{Osborn:2012vt}%
  \BibitemOpen
  \bibfield  {author} {\bibinfo {author} {\bibfnamefont {H.}~\bibnamefont
  {Osborn}},\ }\href {\doibase 10.1016/j.physletb.2012.09.045} {\bibfield
  {journal} {\bibinfo  {journal} {Phys.Lett.}\ }\textbf {\bibinfo {volume}
  {B718}},\ \bibinfo {pages} {169} (\bibinfo {year} {2012})},\ \Eprint
  {http://arxiv.org/abs/1205.1941} {arXiv:1205.1941 [hep-th]} \BibitemShut
  {NoStop}%
%%CITATION = ARXIV:1205.1941;%%
\bibitem [{\citenamefont {Headrick}\ \emph {et~al.}(2015)\citenamefont
  {Headrick}, \citenamefont {Maloney}, \citenamefont {Perlmutter},\ and\
  \citenamefont {Zadeh}}]{Headrick:2015gba}%
  \BibitemOpen
  \bibfield  {author} {\bibinfo {author} {\bibfnamefont {M.}~\bibnamefont
  {Headrick}}, \bibinfo {author} {\bibfnamefont {A.}~\bibnamefont {Maloney}},
  \bibinfo {author} {\bibfnamefont {E.}~\bibnamefont {Perlmutter}}, \ and\
  \bibinfo {author} {\bibfnamefont {I.~G.}\ \bibnamefont {Zadeh}},\ }\href@noop
  {} {\  (\bibinfo {year} {2015})},\ \Eprint {http://arxiv.org/abs/1503.07111}
  {arXiv:1503.07111 [hep-th]} \BibitemShut {NoStop}%
%%CITATION = ARXIV:1503.07111;%%
\bibitem [{\citenamefont {Achucarro}\ and\ \citenamefont
  {Townsend}(1986)}]{Achucarro:1987vz}%
  \BibitemOpen
  \bibfield  {author} {\bibinfo {author} {\bibfnamefont {A.}~\bibnamefont
  {Achucarro}}\ and\ \bibinfo {author} {\bibfnamefont {P.~K.}\ \bibnamefont
  {Townsend}},\ }\href {\doibase 10.1016/0370-2693(86)90140-1} {\bibfield
  {journal} {\bibinfo  {journal} {Phys. Lett.}\ }\textbf {\bibinfo {volume}
  {B180}},\ \bibinfo {pages} {89} (\bibinfo {year} {1986})}\BibitemShut
  {NoStop}%
%%CITATION = PHLTA,B180,89;%%
\bibitem [{\citenamefont {Witten}(1988)}]{Witten:1988hc}%
  \BibitemOpen
  \bibfield  {author} {\bibinfo {author} {\bibfnamefont {E.}~\bibnamefont
  {Witten}},\ }\href {\doibase 10.1016/0550-3213(88)90143-5} {\bibfield
  {journal} {\bibinfo  {journal} {Nucl. Phys.}\ }\textbf {\bibinfo {volume}
  {B311}},\ \bibinfo {pages} {46} (\bibinfo {year} {1988})}\BibitemShut
  {NoStop}%
%%CITATION = NUPHA,B311,46;%%
\bibitem [{\citenamefont {Ba\~nados}(1998)}]{Banados:1998gg}%
  \BibitemOpen
  \bibfield  {author} {\bibinfo {author} {\bibfnamefont {M.}~\bibnamefont
  {Ba\~nados}},\ }\href@noop {} {\  (\bibinfo {year} {1998})},\ \Eprint
  {http://arxiv.org/abs/hep-th/9901148} {arXiv:hep-th/9901148} \BibitemShut
  {NoStop}%
%%CITATION = HEP-TH/9901148;%%
\bibitem [{\citenamefont {Banados}\ and\ \citenamefont
  {Caro}(2004)}]{Banados:2004nr}%
  \BibitemOpen
  \bibfield  {author} {\bibinfo {author} {\bibfnamefont {M.}~\bibnamefont
  {Banados}}\ and\ \bibinfo {author} {\bibfnamefont {R.}~\bibnamefont {Caro}},\
  }\href {\doibase 10.1088/1126-6708/2004/12/036} {\bibfield  {journal}
  {\bibinfo  {journal} {JHEP}\ }\textbf {\bibinfo {volume} {0412}},\ \bibinfo
  {pages} {036} (\bibinfo {year} {2004})},\ \Eprint
  {http://arxiv.org/abs/hep-th/0411060} {arXiv:hep-th/0411060 [hep-th]}
  \BibitemShut {NoStop}%
%%CITATION = HEP-TH/0411060;%%
\bibitem [{\citenamefont {de~Boer}\ and\ \citenamefont
  {Jottar}(2014)}]{deBoer:2014fra}%
  \BibitemOpen
  \bibfield  {author} {\bibinfo {author} {\bibfnamefont {J.}~\bibnamefont
  {de~Boer}}\ and\ \bibinfo {author} {\bibfnamefont {J.~I.}\ \bibnamefont
  {Jottar}},\ }\href@noop {} {\  (\bibinfo {year} {2014})},\ \Eprint
  {http://arxiv.org/abs/1407.3844} {arXiv:1407.3844 [hep-th]} \BibitemShut
  {NoStop}%
%%CITATION = ARXIV:1407.3844;%%
\bibitem [{\citenamefont {Li}\ and\ \citenamefont
  {Theisen}(2015)}]{Li:2015osa}%
  \BibitemOpen
  \bibfield  {author} {\bibinfo {author} {\bibfnamefont {W.}~\bibnamefont
  {Li}}\ and\ \bibinfo {author} {\bibfnamefont {S.}~\bibnamefont {Theisen}},\
  }\href@noop {} {\  (\bibinfo {year} {2015})},\ \Eprint
  {http://arxiv.org/abs/1504.07799} {arXiv:1504.07799 [hep-th]} \BibitemShut
  {NoStop}%
%%CITATION = ARXIV:1504.07799;%%
\bibitem [{\citenamefont {Brown}\ and\ \citenamefont
  {Henneaux}(1986)}]{Brown:1986nw}%
  \BibitemOpen
  \bibfield  {author} {\bibinfo {author} {\bibfnamefont {J.~D.}\ \bibnamefont
  {Brown}}\ and\ \bibinfo {author} {\bibfnamefont {M.}~\bibnamefont
  {Henneaux}},\ }\href@noop {} {\bibfield  {journal} {\bibinfo  {journal}
  {Commun. Math. Phys.}\ }\textbf {\bibinfo {volume} {104}},\ \bibinfo {pages}
  {207} (\bibinfo {year} {1986})}\BibitemShut {NoStop}%
%%CITATION = CMPHA,104,207;%%
\bibitem [{\citenamefont {Ashtekar}\ \emph {et~al.}(1997)\citenamefont
  {Ashtekar}, \citenamefont {Bicak},\ and\ \citenamefont
  {Schmidt}}]{Ashtekar:1996cd}%
  \BibitemOpen
  \bibfield  {author} {\bibinfo {author} {\bibfnamefont {A.}~\bibnamefont
  {Ashtekar}}, \bibinfo {author} {\bibfnamefont {J.}~\bibnamefont {Bicak}}, \
  and\ \bibinfo {author} {\bibfnamefont {B.~G.}\ \bibnamefont {Schmidt}},\
  }\href {\doibase 10.1103/PhysRevD.55.669} {\bibfield  {journal} {\bibinfo
  {journal} {Phys.Rev.}\ }\textbf {\bibinfo {volume} {D55}},\ \bibinfo {pages}
  {669} (\bibinfo {year} {1997})},\ \Eprint
  {http://arxiv.org/abs/gr-qc/9608042} {arXiv:gr-qc/9608042 [gr-qc]}
  \BibitemShut {NoStop}%
%%CITATION = GR-QC/9608042;%%
\bibitem [{\citenamefont {Bagchi}(2011)}]{Bagchi:2010vw}%
  \BibitemOpen
  \bibfield  {author} {\bibinfo {author} {\bibfnamefont {A.}~\bibnamefont
  {Bagchi}},\ }\href {\doibase 10.1007/JHEP02(2011)091} {\bibfield  {journal}
  {\bibinfo  {journal} {JHEP}\ }\textbf {\bibinfo {volume} {1102}},\ \bibinfo
  {pages} {091} (\bibinfo {year} {2011})},\ \Eprint
  {http://arxiv.org/abs/1012.3316} {arXiv:1012.3316 [hep-th]} \BibitemShut
  {NoStop}%
%%CITATION = ARXIV:1012.3316;%%
\bibitem [{\citenamefont {Bagchi}(2013)}]{Bagchi:2013bga}%
  \BibitemOpen
  \bibfield  {author} {\bibinfo {author} {\bibfnamefont {A.}~\bibnamefont
  {Bagchi}},\ }\href {\doibase 10.1007/JHEP05(2013)141} {\bibfield  {journal}
  {\bibinfo  {journal} {JHEP}\ }\textbf {\bibinfo {volume} {1305}},\ \bibinfo
  {pages} {141} (\bibinfo {year} {2013})},\ \Eprint
  {http://arxiv.org/abs/1303.0291} {arXiv:1303.0291 [hep-th]} \BibitemShut
  {NoStop}%
%%CITATION = ARXIV:1303.0291;%%
\bibitem [{\citenamefont {Bagchi}\ and\ \citenamefont
  {Mandal}(2009)}]{Bagchi:2009ca}%
  \BibitemOpen
  \bibfield  {author} {\bibinfo {author} {\bibfnamefont {A.}~\bibnamefont
  {Bagchi}}\ and\ \bibinfo {author} {\bibfnamefont {I.}~\bibnamefont
  {Mandal}},\ }\href {\doibase 10.1016/j.physletb.2009.04.030} {\bibfield
  {journal} {\bibinfo  {journal} {Phys.Lett.}\ }\textbf {\bibinfo {volume}
  {B675}},\ \bibinfo {pages} {393} (\bibinfo {year} {2009})},\ \Eprint
  {http://arxiv.org/abs/0903.4524} {arXiv:0903.4524 [hep-th]} \BibitemShut
  {NoStop}%
%%CITATION = ARXIV:0903.4524;%%
\bibitem [{\citenamefont {Bagchi}\ \emph {et~al.}(2010)\citenamefont {Bagchi},
  \citenamefont {Gopakumar}, \citenamefont {Mandal},\ and\ \citenamefont
  {Miwa}}]{Bagchi:2009pe}%
  \BibitemOpen
  \bibfield  {author} {\bibinfo {author} {\bibfnamefont {A.}~\bibnamefont
  {Bagchi}}, \bibinfo {author} {\bibfnamefont {R.}~\bibnamefont {Gopakumar}},
  \bibinfo {author} {\bibfnamefont {I.}~\bibnamefont {Mandal}}, \ and\ \bibinfo
  {author} {\bibfnamefont {A.}~\bibnamefont {Miwa}},\ }\href {\doibase
  10.1007/JHEP08(2010)004} {\bibfield  {journal} {\bibinfo  {journal} {JHEP}\
  }\textbf {\bibinfo {volume} {1008}},\ \bibinfo {pages} {004} (\bibinfo {year}
  {2010})},\ \Eprint {http://arxiv.org/abs/0912.1090} {arXiv:0912.1090
  [hep-th]} \BibitemShut {NoStop}%
%%CITATION = ARXIV:0912.1090;%%
\bibitem [{\citenamefont {Deser}\ \emph {et~al.}(1982)\citenamefont {Deser},
  \citenamefont {Jackiw},\ and\ \citenamefont {Templeton}}]{Deser:1982vy}%
  \BibitemOpen
  \bibfield  {author} {\bibinfo {author} {\bibfnamefont {S.}~\bibnamefont
  {Deser}}, \bibinfo {author} {\bibfnamefont {R.}~\bibnamefont {Jackiw}}, \
  and\ \bibinfo {author} {\bibfnamefont {S.}~\bibnamefont {Templeton}},\ }\href
  {\doibase 10.1103/PhysRevLett.48.975} {\bibfield  {journal} {\bibinfo
  {journal} {Phys. Rev. Lett.}\ }\textbf {\bibinfo {volume} {48}},\ \bibinfo
  {pages} {975} (\bibinfo {year} {1982})}\BibitemShut {NoStop}%
%%CITATION = PRLTA,48,975;%%
\bibitem [{\citenamefont {Deser}\ and\ \citenamefont
  {Kay}(1983)}]{Deser:1982sw}%
  \BibitemOpen
  \bibfield  {author} {\bibinfo {author} {\bibfnamefont {S.}~\bibnamefont
  {Deser}}\ and\ \bibinfo {author} {\bibfnamefont {J.~H.}\ \bibnamefont
  {Kay}},\ }\href@noop {} {\bibfield  {journal} {\bibinfo  {journal} {Phys.
  Lett.}\ }\textbf {\bibinfo {volume} {B120}},\ \bibinfo {pages} {97} (\bibinfo
  {year} {1983})}\BibitemShut {NoStop}%
%%CITATION = PHLTA,B120,97;%%
\bibitem [{\citenamefont {Bergshoeff}\ \emph {et~al.}(2009)\citenamefont
  {Bergshoeff}, \citenamefont {Hohm},\ and\ \citenamefont
  {Townsend}}]{Bergshoeff:2009hq}%
  \BibitemOpen
  \bibfield  {author} {\bibinfo {author} {\bibfnamefont {E.~A.}\ \bibnamefont
  {Bergshoeff}}, \bibinfo {author} {\bibfnamefont {O.}~\bibnamefont {Hohm}}, \
  and\ \bibinfo {author} {\bibfnamefont {P.~K.}\ \bibnamefont {Townsend}},\
  }\href {\doibase 10.1103/PhysRevLett.102.201301} {\bibfield  {journal}
  {\bibinfo  {journal} {Phys. Rev. Lett.}\ }\textbf {\bibinfo {volume} {102}},\
  \bibinfo {pages} {201301} (\bibinfo {year} {2009})},\ \Eprint
  {http://arxiv.org/abs/0901.1766} {arXiv:0901.1766 [hep-th]} \BibitemShut
  {NoStop}%
%%CITATION = 0901.1766;%%
\bibitem [{\citenamefont {Andringa}\ \emph {et~al.}(2010)\citenamefont
  {Andringa} \emph {et~al.}}]{Andringa:2009yc}%
  \BibitemOpen
  \bibfield  {author} {\bibinfo {author} {\bibfnamefont {R.}~\bibnamefont
  {Andringa}} \emph {et~al.},\ }\href {\doibase 10.1088/0264-9381/27/2/025010}
  {\bibfield  {journal} {\bibinfo  {journal} {Class. Quant. Grav.}\ }\textbf
  {\bibinfo {volume} {27}},\ \bibinfo {pages} {025010} (\bibinfo {year}
  {2010})},\ \Eprint {http://arxiv.org/abs/0907.4658} {arXiv:0907.4658
  [hep-th]} \BibitemShut {NoStop}%
%%CITATION = 0907.4658;%%
\bibitem [{\citenamefont {Bergshoeff}\ \emph {et~al.}(2013)\citenamefont
  {Bergshoeff}, \citenamefont {de~Haan}, \citenamefont {Hohm}, \citenamefont
  {Merbis},\ and\ \citenamefont {Townsend}}]{Bergshoeff:2013xma}%
  \BibitemOpen
  \bibfield  {author} {\bibinfo {author} {\bibfnamefont {E.~A.}\ \bibnamefont
  {Bergshoeff}}, \bibinfo {author} {\bibfnamefont {S.}~\bibnamefont {de~Haan}},
  \bibinfo {author} {\bibfnamefont {O.}~\bibnamefont {Hohm}}, \bibinfo {author}
  {\bibfnamefont {W.}~\bibnamefont {Merbis}}, \ and\ \bibinfo {author}
  {\bibfnamefont {P.~K.}\ \bibnamefont {Townsend}},\ }\href {\doibase
  10.1103/PhysRevLett.111.111102, 10.1103/PhysRevLett.111.259902} {\bibfield
  {journal} {\bibinfo  {journal} {Phys.Rev.Lett.}\ }\textbf {\bibinfo {volume}
  {111}},\ \bibinfo {pages} {111102} (\bibinfo {year} {2013})},\ \Eprint
  {http://arxiv.org/abs/1307.2774} {arXiv:1307.2774} \BibitemShut {NoStop}%
%%CITATION = ARXIV:1307.2774;%%
\bibitem [{\citenamefont {Henneaux}\ and\ \citenamefont
  {Rey}(2010)}]{Henneaux:2010xg}%
  \BibitemOpen
  \bibfield  {author} {\bibinfo {author} {\bibfnamefont {M.}~\bibnamefont
  {Henneaux}}\ and\ \bibinfo {author} {\bibfnamefont {S.-J.}\ \bibnamefont
  {Rey}},\ }\href {\doibase 10.1007/JHEP12(2010)007} {\bibfield  {journal}
  {\bibinfo  {journal} {JHEP}\ }\textbf {\bibinfo {volume} {1012}},\ \bibinfo
  {pages} {007} (\bibinfo {year} {2010})},\ \Eprint
  {http://arxiv.org/abs/1008.4579} {arXiv:1008.4579 [hep-th]} \BibitemShut
  {NoStop}%
\bibitem [{\citenamefont {Campoleoni}\ \emph {et~al.}(2010)\citenamefont
  {Campoleoni}, \citenamefont {Fredenhagen}, \citenamefont {Pfenninger},\ and\
  \citenamefont {Theisen}}]{Campoleoni:2010zq}%
  \BibitemOpen
  \bibfield  {author} {\bibinfo {author} {\bibfnamefont {A.}~\bibnamefont
  {Campoleoni}}, \bibinfo {author} {\bibfnamefont {S.}~\bibnamefont
  {Fredenhagen}}, \bibinfo {author} {\bibfnamefont {S.}~\bibnamefont
  {Pfenninger}}, \ and\ \bibinfo {author} {\bibfnamefont {S.}~\bibnamefont
  {Theisen}},\ }\href {\doibase 10.1007/JHEP11(2010)007} {\bibfield  {journal}
  {\bibinfo  {journal} {JHEP}\ }\textbf {\bibinfo {volume} {1011}},\ \bibinfo
  {pages} {007} (\bibinfo {year} {2010})},\ \Eprint
  {http://arxiv.org/abs/1008.4744} {arXiv:1008.4744 [hep-th]} \BibitemShut
  {NoStop}%
\bibitem [{\citenamefont {Gary}\ \emph {et~al.}(2012)\citenamefont {Gary},
  \citenamefont {Grumiller},\ and\ \citenamefont {Rashkov}}]{Gary:2012ms}%
  \BibitemOpen
  \bibfield  {author} {\bibinfo {author} {\bibfnamefont {M.}~\bibnamefont
  {Gary}}, \bibinfo {author} {\bibfnamefont {D.}~\bibnamefont {Grumiller}}, \
  and\ \bibinfo {author} {\bibfnamefont {R.}~\bibnamefont {Rashkov}},\ }\href
  {\doibase 10.1007/JHEP03(2012)022} {\bibfield  {journal} {\bibinfo  {journal}
  {JHEP}\ }\textbf {\bibinfo {volume} {1203}},\ \bibinfo {pages} {022}
  (\bibinfo {year} {2012})},\ \Eprint {http://arxiv.org/abs/1201.0013}
  {arXiv:1201.0013 [hep-th]} \BibitemShut {NoStop}%
%%CITATION = ARXIV:1201.0013;%%
\bibitem [{\citenamefont {Henneaux}\ \emph {et~al.}(2012)\citenamefont
  {Henneaux}, \citenamefont {Lucena~G{\'o}mez}, \citenamefont {Park},\ and\
  \citenamefont {Rey}}]{Henneaux:2012ny}%
  \BibitemOpen
  \bibfield  {author} {\bibinfo {author} {\bibfnamefont {M.}~\bibnamefont
  {Henneaux}}, \bibinfo {author} {\bibfnamefont {G.}~\bibnamefont
  {Lucena~G{\'o}mez}}, \bibinfo {author} {\bibfnamefont {J.}~\bibnamefont
  {Park}}, \ and\ \bibinfo {author} {\bibfnamefont {S.-J.}\ \bibnamefont
  {Rey}},\ }\href {\doibase 10.1007/JHEP06(2012)037} {\bibfield  {journal}
  {\bibinfo  {journal} {JHEP}\ }\textbf {\bibinfo {volume} {1206}},\ \bibinfo
  {pages} {037} (\bibinfo {year} {2012})},\ \Eprint
  {http://arxiv.org/abs/1203.5152} {arXiv:1203.5152 [hep-th]} \BibitemShut
  {NoStop}%
%%CITATION = ARXIV:1203.5152;%%
\end{thebibliography}

%

\end{document}